\documentclass[twocolumn]{aastex631}

\newcommand{\gammaray}{\mbox{Gamma-ray}}

\usepackage{amsmath,amsfonts,amssymb,amsthm}

\usepackage{booktabs}
\usepackage{array}
\usepackage{tabularx}
\usepackage{makecell}
\usepackage{enumitem}
\usepackage{hyperref}
\usepackage{cleveref}
\usepackage{soul}

\begin{document}

\title{Interpreting the multi-TeV emission from GRB~221009A with a second electron component accelerated by turbulence in the jet}

\author{Xing-Wei Gong}
\affiliation{School of Astronomy and Space Science, Nanjing University, Xianlin Avenue 163, Nanjing 210023, China}

\author[0000-0003-1576-0961]{Ruo-Yu Liu}
\affiliation{School of Astronomy and Space Science, Nanjing University, Xianlin Avenue 163, Nanjing 210023, China}
\affiliation{Key laboratory of Modern Astronomy and Astrophysics (Nanjing University), Ministry of Education, China}
\affiliation{Tianfu Cosmic Ray Research Center,
 Chengdu 610000, Sichuan, China}

\author[0000-0003-4621-0807]{Ze-Lin Zhang}
\affiliation{School of Electrical and Electronic Engineering, Anhui Science and Technology University,\\Huangshan Avenue 1501, Bengbu 233030, China}

\author[0000-0001-9064-160X]{Katsuaki Asano}
\affiliation{Institute for Cosmic Ray Research, The University of Tokyo, 5-1-5 Kashiwanoha, Kashiwa, Chiba 277-8582, Japan}

\author[0000-0002-2395-7812]{Martin Lemoine}
\affiliation{Astroparticule et Cosmologie (APC), CNRS - Université Paris Cité, 75013 Paris, France}
\correspondingauthor{Ruo-Yu Liu}
\email{ryliu@nju.edu.cn}

\begin{abstract}
    The detection of very high-energy (VHE) afterglow emission of the gamma-ray burst (GRB)~221009A by the Large High Altitude Air Shower Observatory (LHAASO) provides a unique opportunity to probe particle acceleration in relativistic outflows. The hard spectrum at multi-TeV band cannot be fully explained by synchrotron-self-Compton radiation of the conventional one-zone afterglow model. In this work, we introduce a second component of relativistic electrons from stochastic acceleration via downstream turbulence of the external shock. Using a Fokker-Planck approach to model the evolution of protons and electrons, and the non-linear feedback of turbulence damping, we show that the inverse Compton radiation of the second electron component may harden the observed spectrum above multi-TeV energy, and significantly ameliorate the fitting to the spectral evolution measured by LHAASO without violating lower-energy observations. We also discuss the potential presence of the second electron component in other GRB afterglows, which may provide a possible observational signature for future studies.

\end{abstract}


\section{Introduction} \label{sec:intro}

Gamma-ray bursts (GRBs) are extremely powerful events that produce bright gamma-ray radiation ranging from keV to sub-MeV at the prompt outburst.
The afterglow phase follows the prompt emission, producing broad-band emission from radio to gamma rays, which can last days to months.
From a theoretical view, afterglow emission is produced when ejecta from central engine interacts with the circumburst medium  \citep{Rees1992MNRAS,Meszaros1993ApJ} (see \citet{Miceli2022Galax} for a review).
Phenomenological models are helpful to explain the observational features in the afterglow.
In the standard afterglow model, upstream interstellar medium (ISM) particles being swept into the forward shock of the afterglow form a non-thermal power-law distribution due to diffusive shock acceleration.
These electrons contribute to the spectral energy distribution (SED) from radio to hard X-rays  via synchrotron radiation. Under extreme conditions, synchrotron emission may extend up to GeV energies \citep{Kumar2009MNRAS,Kumar2010MNRAS}. Emission in the multi-GeV energies and very-high energy (VHE, $E > 0.1$~ TeV) band, is dominated by the inverse Compton (IC) emission, and in most cases the synchrotron-self-Compton (SSC) emission \citep{Sari2001ApJ,Zhang2001ApJ,Wang2001ApJ,Zou2009MNRAS}.

Several GRBs have been detected in the VHE band in recent years, e.g. GRB~190114C \citep{MAGICCollaboration2019aNatur,MAGICCollaboration2019bNatur}, GRB~180720B \citep{Abdalla2019Natur}, GRB~190829A\citep{H.E.S.S.Collaboration2021Sci}, and GRB~201216C \citep{Abe2024MNRAS}, where the VHE emission is interpreted to originate from the SSC component of the forward shock.
On 9 October 2022 at 13:16:59.99 UTC, GRB~221009A triggered the Fermi \gammaray burst monitor (Fermi GBM) \citep{Lesage2023ApJL}.
GRB~221009A is referred to as the brightest of all time gamma-ray burst due to its intrinsic brightness ($E_0\sim1.5\times10^{51}\,\rm{erg}$) and unusual proximity ($z=0.151$) \citep{Burns2023ApJL}.
The Large High Altitude Air Shower Observatory (LHAASO) recorded more than 64,000 Photons at 0.2$-$7 \rm{TeV} between 0$-$3000~s after trigger with the Water Cherenkov Detector Array (WCDA) \citep{LHAASOCollaboration2023Sci}, and further reported detection of 140 gamma rays above 3 TeV at 230$-$900~s after trigger by the Kilometer Square Array (KM2A) \citep{Cao2023SciA}.
This is the first time that both the prompt and afterglow phases of a TeV-detected GRB have been covered by observation \citep{LHAASOCollaboration2023Sci}.
The recorded spectrum at afterglow phase cannot be explained with the one-zone SSC model \citep{Cao2023SciA}.
The onset of afterglow at $T^*$ is 226~s after trigger time $T_0$.
At $T^*+100$~s, the observed spectra at 0.3$-$5 TeV by LHAASO-WCDA becomes harder than predicted by the one-zone SSC model.
Therefore, a second particle component is required to account  for the hard spectrum.

Recent works claim a structured two-zone jet model of GRB~221009A, including a Poynting-flux-dominated core and matter-dominated wing \citep{Zheng2024ApJ,Zhang2024JHEAp,Ren2024ApJ}.
\cite{Zhang2025JHEAp} improved the fit of the evolution of the spectral index at TeV band by adding a component of the reverse shock proton synchrotron emission. 
In this letter, we propose another explanation with the non-thermal electrons by stochastic acceleration via turbulence in the downstream of the forward shock.

Stochastic particle acceleration via turbulence has long been considered an effective mechanism of cosmic ray acceleration, and has been applied to GRB afterglows and other astrophysical conditions, e.g. \citet{Asano2016PhRvD,Zhang2021PhRvD,Sciaccaluga2022MNRAS,Dermer1996ApJ,Liu2006ApJ,Mertsch2011PhRvL,Kakuwa2016ApJ}. {This energization process is conveniently described by diffusion in energy space, as in the original picture proposed by E. Fermi \citep{Fermi1949PhRv}, and thus characterized by a diffusion coefficient $D_{EE}$.  In the standard framework of quasi-linear theory (QLT),} the  diffusion coefficient is calculated through linear perturbation theory, and through the averaging over an ensemble of pre-existing weak turbulent waves \citep{Schlickeiser1989ApJ,Schlickeiser1993JPlPh,Jaekel1992JPhG}. Correspondingly, it is usually assumed that the fluctuating turbulent magnetic field is smaller than the background average magnetic field, i.e. $\delta B < B_0$, {and that particle energization occurs through wave-particle interactions.}

Under this weak field condition, turbulent magnetic fluctuation are decomposed into non-interacting fast and slow magnetosonic waves, and incompressible Alfv\'{e}n modes \citep{Kulsrud2005ppa}. {Wave-particle interactions are dominated by resonances of the form }
 $k_\parallel v_\parallel  - \omega_\mathrm{w} + n\Omega = 0$, where $k_\parallel$ denotes the parallel component of wave-vector with respect to the mean magnetic field, $v$ is the particle velocity, $\omega_\mathrm{w}$ refers to the wave frequency, and $\Omega=eB/mc$ the gyro-frequency of particles.
The particle is in gyroresonance for $n\neq 0$, and transit-time damping (TTD) for $n=0$.

\cite{Zhang2021PhRvD} explored the stochastic acceleration of protons in a GRB afterglow. In this setting, the pre-existing circumburst magnetic field is amplified by compression and microscopic current filamentation instabilities that develop in the shock precursor, to form the downstream turbulence  \citep{Sironi2013ApJ,Garasev2016MNRAS,Lemoine2019PhRvL,Groselj2024ApJL}. While these magnetic structures appear essentially static and thus not prone to accelerate particles, the resulting turbulence can nevertheless undergo nonlinear processing and be further amplified by macroscopic MHD instabilities. For instance, the Rayleigh-Taylor instability at the contact discontinuity \citep{Levinson2009ApJL} and the Richtmyer-Meshkov instability arising from the interaction between the shock and upstream density inhomogeneities \citep{Inoue2011ApJ} provide  natural reservoirs of turbulence where particles can undergo acceleration.
\cite{Zhang2021PhRvD} utilized a Fokker-Planck approach of modeling the particle acceleration and non-linear feedback of turbulence damping. While Alfv\'{e}n and slow modes preferentially cascade perpendicular to the magnetic field, the compressible fast mode shows isotropic cascade and therefore serves as the dominant cosmic ray scatterer \citep{Yan2002PhRvL, Makwana2020PhRvX}. In \cite{Zhang2021PhRvD}, acceleration was due to particles in gyro-resonance with fast mode wave.

In this Letter, we investigate the acceleration of electrons and ions by both gyro-resonance and TTD with fast mode wave based on their theoretical framework. Using this model, we provide a self-consistent explanation of the VHE observations by LHAASO and observations in other wavelengths. The sections are organized as follows.
In Section~\ref{sec:model}, we describe the physical model for stochastic acceleration.
In Section~\ref{sec:Results and Discussion}, we display and discuss the results.
Finally, we summarize our findings and give conclusions in Section~\ref{sec:summary}.

\section{Model Description}\label{sec:model}
The standard afterglow model offers a satisfactory explanation of the early VHE gamma-ray afterglow via the SSC of electrons accelerated at the shock (hereafter, we call it ``SH electrons'' for simplicity), except for the SED at $\lesssim 10\,$TeV~\citep{Cao2023SciA}. In addition to the SH electrons, we modify the model by introducing a second electron component that is accelerated in the downstream turbulence via the stochastic acceleration (hereafter, we call it ``SA electrons'' for simplicity). The modeling of SH particles follows the standard one-zone SSC model of afterglow, as \cite{LHAASOCollaboration2023Sci}. To compute the spectral evolution of the SA electrons, we use the same theoretical framework as \cite{Zhang2021PhRvD} by solving two coupled Fokker-Planck equations describing the evolutions of the particle spectrum and of the turbulence spectrum, respectively. 

\subsection{Stochastic Acceleration \label{sec:SA}}
Our model involves two zones. SH electrons are accelerated by the shock in the first zone at shock front, as already studied by \citep{Cao2023SciA}.   
The stochastic acceleration takes place in the second zone located in the downstream of the GRB external shock, where turbulence is excited to substantial levels by MHD instabilities. The acceleration efficiency and radiation of accelerated particles then depend on the properties of the downstream plasma. We therefore need to first obtain the dynamical evolution of an adiabatic afterglow, which could be described with \cite{Huang1999MNRAS}:
\begin{equation}
    \frac{{\rm d}\Gamma}{{\rm d} m} \simeq -\frac{\Gamma^2-1}{M_{\mathrm{ej}} + 2\Gamma m}\text{,}\label{dynevo}
\end{equation}
where $\Gamma$ is the bulk Lorentz factor of the shocked ISM, $m$ is the rest mass of the ISM within the shock, and $M_{\mathrm{ej}}$ is the mass of the GRB ejecta.

Conventionally, we use a Fokker-Planck description to follow the spectrum of particles accelerated by turbulence \citep{Miller1996ApJ},
\begin{equation}
    \begin{split}
        \frac{\partial N}{\partial t}=\,&\frac{\partial }{\partial E}\left[D_{{EE}}(E,t) \frac{\partial N}{\partial E}\right] -  \frac{\partial }{\partial E}\left[\frac{2D_{{EE}}(E,t)}{E}N\right]\\ &
        -\frac{\partial }{\partial E}\left(\langle \dot{{E}} \rangle N\right) - \frac{N}{t_{\mathrm{esc}}} + Q_{\mathrm{inj}}(E,t)\text{.}
    \end{split}\label{particle pde}
\end{equation}
The time $t$ here is measured in the comoving frame of the downstream plasma, and we let $t=0$ at the onset of afterglow. The time in observer frame is denoted with $T$. These two times are connected with the relation $T-T^* = t/2\Gamma$.
In the first term on the right-hand side (r.h.s.) of the equation, 
\begin{equation}
    \begin{split}
        D_{EE}(E) = \frac{E^2 \beta_\mathrm{ph}^2 c k_\mathrm{res}^2}{U_B}\left[ \int_{k_\mathrm{res}} \frac{W_\mathrm{F}(k)}{k}\, \mathrm{d}k\right.\\
        \left. + \int \Phi(E,k) W_\mathrm{F} (k) \, \mathrm{d}k\right]
    \end{split}\label{dee}
\end{equation}  
is the diffusion coefficient of particles in the energy space, where $\beta_\mathrm{ph} = v_\mathrm{ph}/c$ is the dimensionless phase velocity for fast mode (see \Cref{app:kernel}), and $k_\mathrm{res} = eB/E$ is the resonance wavenumber of turbulence with a particle of energy $E$.
The first term on the r.h.s. denotes the contribution of gyro-resonance \citep{Kakuwa2016ApJ}, and the second term represents TTD. 
$\Phi(E,k)$ here is the resonance kernel adopted from \cite{Lemoine2024PhRvD} (see \Cref{app:kernel}). $U_B$ is the local magnetic energy density. $W_\mathrm{F}$ is the fast mode magnetic component of downstream fluctuation, which is computed by $W_\mathrm{F}(k) = \alpha_\mathrm{F}  W(k)$, with $W(k)$ the power spectrum of total turbulent energy and $\alpha_\mathrm{F}$ is a dimensionless factor. 
The escape timescale $t_{\mathrm{esc}} \equiv (R/ \Gamma v_\mathrm{ph} )^2/t_{acc}$ \citep{Tramacere2011ApJ} denotes the typical timescale a particle escapes the acceleration zone, where $t_\mathrm{acc} \equiv E^2 / D_{EE}$ is the timescale for acceleration. The above formulation of particle escape follows a leaky box treatment; however, it turns out that escape plays an unimportant role. 
The $\langle\dot{E}\rangle$ in the second term on the r.h.s. of Equation~(\ref{particle pde}) denotes the particle energy loss rate.
For electrons, radiation loss must be included in addition to adiabatic loss.

The particle injection term $Q_\mathrm{inj}$ comes from the shock front. Nonthermal SH particles are injected into the SA region and then transported further downstream by advection, where the turbulent energy density is lower because of wave decay or damping. We assume that once SH particles are advected away from the shock front, they cannot return, and a fraction $f_{\mathrm{ra}}$ of them get involved in the stochastic acceleration in downstream turbulence (i.e., become SA particles). 
SH particles already form a non-thermal power-law distribution before being transported away from the shock front and being accelerated by the downstream turbulence. 
Thus, the injection term $Q_\mathrm{inj}  (E,t) = f_{\mathrm{ra}} N_\mathrm{sh} (E,t)/t_\mathrm{adv}$, where $N_{\rm sh}(E,t)$ is the spectrum of SH particles predicted by the standard afterglow model, and $t_\mathrm{adv}$ is the advection timescale for SH particles to leave the shock front, which is $3R/\Gamma c$ in the ultra-relativistic limit \citep{Zhang2019pgrb}.

\begin{figure}[ht]
    \centering
    \includegraphics[width=\linewidth]{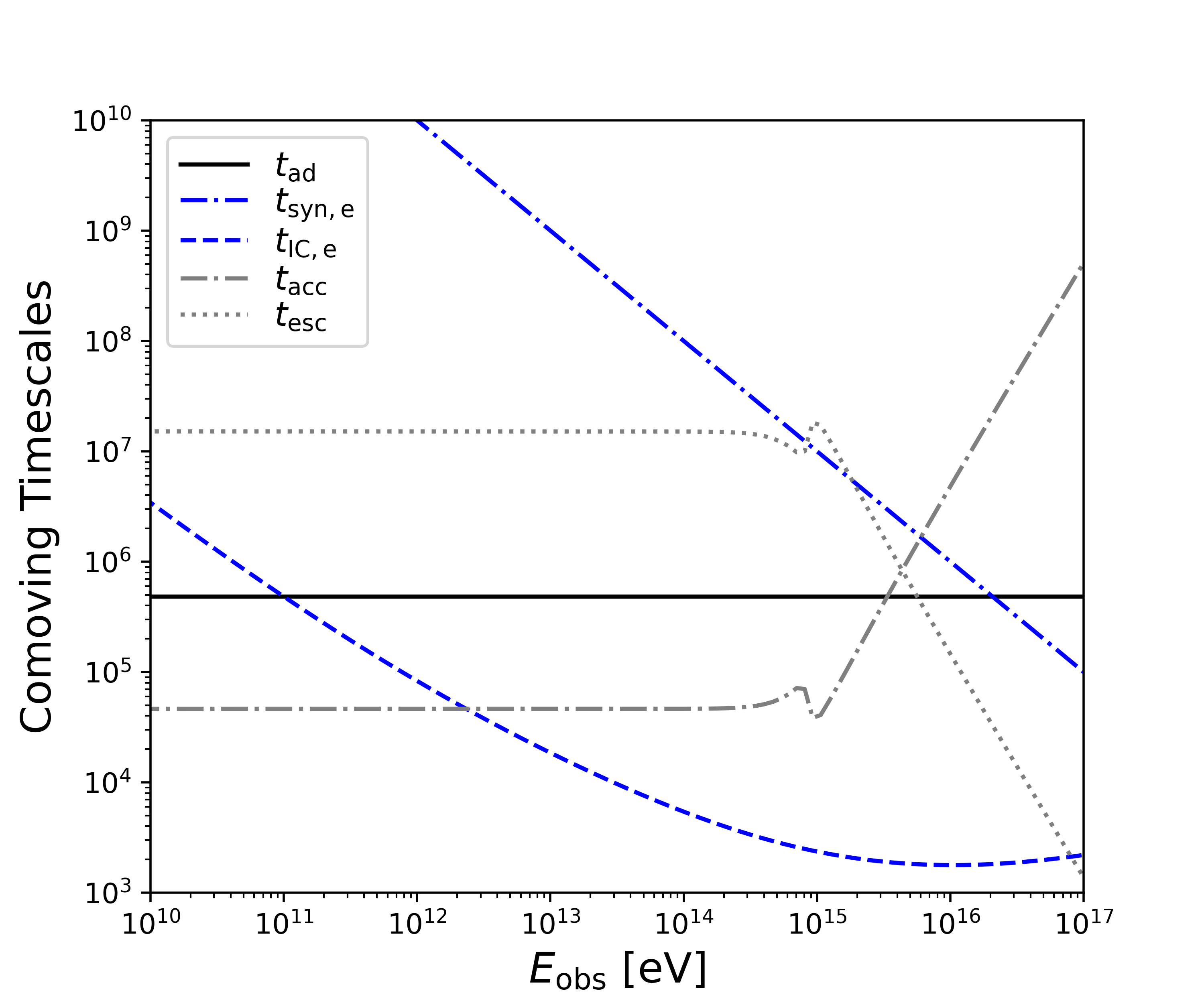}
    \caption{The evolution timescales at $T^* + 1000$~s, with parameters in \Cref{para}. The black solid line indicates the dynamical timescale; the gray dash-dotted line is the acceleration timescale; the gray dotted line is the escape timescale. These timescales are applicable to both electrons and protons. The radiation cooling timescales for SA electrons are plotted in blue lines, where dashed line is the IC cooling timescale, and  the dash-dotted line is the synchrotron cooling timescale. All timescales are measured in the comoving frame. For definitions of relevant timescales, see \Cref{app:timescale}.}
    \label{fig:timescales}
\end{figure}

Numerical simulations indicate that electrons enter the shock with energies nearly equal to ions due to strong pre-shock heating mechanisms, and hence experience similar acceleration processes \citep{Sironi2013ApJ,2022ApJ...930L...8V}. 
As a result, the injection spectra of protons and electrons in our model are non-thermal power-law spectra with the same spectral index and comparable minimum energies. 
However, SH protons differ from electrons by not being affected by radiation cooling.
This is because the radiation of protons, via either the synchrotron/IC processes, or hadronic interactions such as the $pp$ and $p\gamma$ processes, are unimportant for the parameters considered here.  
As shown in \Cref{fig:timescales}, 
the electron cooling timescale is depicted in blue dash-dotted line. As the proton radiation power is weaker than electron by a factor of $(m_p/m_e)^4\sim 10^{13}$, the proton radiation cooling timescale is much longer than the dynamical timescale and therefore the proton cooling can be safely ignored. At $T^* + 1000$~s, the SA proton peak energy is around $10$ TeV in comoving frame. The photon number density in $\Delta$-resonance ($\epsilon_\mathrm{ph} \sim 0.3$ MeV) with these protons is approximately $6\times 10^{-4} \,\text{cm}^{-3}$, so the optical depth of $p\gamma$ process is $\tau  = n_\mathrm{ph}\sigma_{p\gamma} R/\Gamma \sim 5\times 10^{-7} \ll 1$.

\subsection{Turbulence Evolution}
Turbulence in the wave number $k-$space can be modeled with $W(k) \propto k ^{-q}$, either with $q=5/3$ proposed by \cite{Kolmogorov1941DoSSR}, hereafter K41, or $q = 3/2$ by \cite{Kraichnan1965PhFl} and \cite{Iroshnikov1963AZh}, hereafter IK. The cascade of the turbulence in the wave number $k-$space can be described with a Fokker-Planck equation \citep{Miller1996ApJ}
\begin{equation}
    \begin{split}
        \frac{\partial W}{\partial t} = \frac{\partial }{\partial k}\left[D_{\mathrm{kk}}(k,t) \frac{\partial W}{\partial k}\right] - \frac{\partial }{\partial k}\left[\frac{2D_{\mathrm{kk}}(k,t)}{k}W\right] \\
        + \frac{k}{3}(\nabla\cdot \mathrm{v}) \frac{\partial W}{\partial k}+ \Gamma_{\mathrm{w}}(k,t) W + Q_{\mathrm{w,inj}}(k,t)\text{,}
    \end{split}\label{turb pde}
\end{equation}
where $\Gamma_{\rm w}$ is the damping rate due to particle acceleration (see \Cref{app:damp} for a brief derivation). It reflects the energy loss of turbulence as the energy is transferred from waves to particles. $Q_{\rm w,inj}$ is the turbulence injection rate, and $D_{\mathrm{kk}}$ denotes the turbulence diffusion coefficient in the wave number space \citep{Zhou1990JGR,Miller1996ApJ},
\begin{equation}
    D_{\mathrm{kk}}(k) = \left\{\begin{aligned}
         & v_{\mathrm{ph}} k^{7/2}\left[\frac{W(k)}{2U_B}\right]^{1/2} & \rm{(K41)} \\
         & v_{\mathrm{ph}} k^{4}\left[\frac{W(k)}{2U_B}\right]         & \rm{(IK)}
    \end{aligned}\right.\textbf{,}
\end{equation}
The third term on the r.h.s. of \Cref{turb pde} denotes the adiabatic energy loss due to the expansion of the external shock.

We further assume a factor $\alpha$ to describe the magnetic component of turbulence, i.e. $W_B(k) = \alpha W(k)$. Beside the fast magnetosonic wave mode, the contribution from Alfv\'{e}n and slow modes are included within $W_B$, despite their inefficiency in cosmic ray acceleration due to the anisotropic cascade \citep{Chandran2000PhRvL,Cho2002ApJ,Cho2003MNRAS,Yan2002PhRvL,Yan2004ApJ}.  Therefore, the total magnetic energy density is $U_B = U_{B0}+\int \alpha W(k)\, \mathrm{d}k$, where $U_{B0}$ denotes the energy density of the regular magnetic field that does not decay with damping of turbulence, and the second term on the r.h.s. represents the energy density of the turbulent magnetic field. With the assumption that the downstream plasma is highly turbulent, the energy conservation requires $\alpha\varepsilon_T \approx \varepsilon_B$, where $\varepsilon_T$ is the equipartition coefficient of the total turbulent energy. It denotes the fraction of turbulent energy in the total energy of the shocked fluid. The $\varepsilon_B$ is the magnetic equipartition factor of downstream plasma, which will decay as the plasma leaves the shock front \citep{Lemoine2013MNRAS}, but then further amplified by MHD instabilities and thus can be larger than $\varepsilon_{B,\mathrm{sh}}$ immediately behind the shock.

As the MHD instability amplifies the the magnetic field in the shocked plasma,  we assume the turbulence injection into the SA region is related to the energy flux at the shock front. Specifically, the turbulent injection is tied to the energy swept into the shock via the equipartition factor $\varepsilon_T$. 
Therefore, the injection rate of the turbulence is modeled with a $\delta$-like function $Q_{\rm w, inj}(k) = 4 \Gamma^2 \varepsilon_T n_{\mathrm{ISM}} m_\mathrm{p} c^2 / (R/\Gamma c) \, \delta(k-k_\mathrm{inj} )$. 
The $k_\mathrm{inj}$ here is the minimum wave number of the inertial range.
We define $\xi = \lambda_\mathrm{inj}   / (R/\Gamma)$ as the dimensionless eddy scale at which turbulence is injected.
The injection scale $\xi$ is uncertain, thus we treat it as a free parameter, as long as the wavelength of turbulence injection is much larger than the skin depth of the upstream plasma ${(\Gamma m_\mathrm{p}c^2/4 \pi n_{\mathrm{ISM}}e^2)}^{1/2}$.

\subsection{Particle Radiation \label{radiation}}

The radiation comes mainly from relativistic electrons.
We calculate their synchrotron radiation with the total mean magnetic field $\overline{B} = \sqrt{8\pi  U_B}$, and the IC scattering including the Klein-Nishina effect which softens the spectrum at high energies. 
The radiative cooling of SA electrons includes the synchrotron cooling, SSC cooling and the IC scatter off the synchrotron radiation of SH electrons. The radiation of SA electrons, on the other hand, is not considered as the target field of IC of SH electrons,  because the radiation of SA electrons will be  only important at VHE band, as will be shown later. 
As mentioned in the above section, the two zones in our model, SH and SA, have different magnetization. The electron synchrotron radiation in the two regions is therefore calculated with different magnetic equipartition factors $\varepsilon_{B,\mathrm{sh}}$ and $\varepsilon_{B}$, respectively. 

For photons of energy above TeV, pair production becomes important.
We calculate the intrinsic $\gamma\gamma$ absorption by soft photons from synchrotron radiation of electrons at the shock and in the downstream turbulence (\Cref{app:radiation}). 

\subsection{Parameters Setup}

\begin{table*}[ht]
    \centering
    \begin{tabular}{ccc}
        \toprule
        Symbol              & Definition                     & Value                    \\
        \midrule

        $\Theta$            & {jet opening angle }             & 0.8$^\circ$              \\
        $\varepsilon_{B,\mathrm{sh}}$ &{magnetic equipartititon factor (shock)}& $1.5\times 10^{-3}$      \\
        $\varepsilon_B$     &  {magnetic equipartition factor (downstream)} & $4\times 10^{-3}$      \\
        $\varepsilon_T$     &  {turbulent equipartition factor} & $1\times 10^{-2}$     \\
        $\alpha$            &  {turbulence magnetic component}  & $ 0.4$                   \\
        $\alpha_\mathrm{F} $          &  {fast mode magnetic component}   & $0.1$                    \\
        $\mathcal{E}_\mathrm{tot}$ &  {isotropic energy}               & $1.5 \times 10^{55}$ erg \\
        $p$                 &  {spectral index}                 & 2.22                     \\
        $z$                 &  {redshift}                       & 0.1505                   \\
        $\varepsilon_\mathrm{e}$     &  {electron equipartition factor}  & 0.025                    \\
        $n_{\mathrm{ISM}}$  &  {interstellar medium density}    & 0.26 cm$^{-3}$           \\
        $\Gamma_0$          &  {initial bulk Lorentz factor}    & $580$                    \\
        $\xi$               &  {turbulence scale (in $R/\Gamma$)} & $0.01$                      \\
        $R_0$               &  {initial shock radius}           & $1.63\times 10^{15}$ cm  \\
        $f_{\mathrm{ra}}$  &  {re-accelerated particle fraction}  & $2\times 10^{-3}$      \\
        $B_{\mathrm{ISM}}$  &  {upstream magnetic field}        & $5$ $\mu$G               \\
        \bottomrule
    \end{tabular}
    \caption{Key parameters describing our two-component model.}
    \label{para}
\end{table*}

We found that the initial state of the magnetic field and the turbulence spectrum play a minor role in the calculation, so we simply set the initial turbulence energy density $W(k) = 0$, thus $B_0= 0$ at $T=T^*$.
The employ $\alpha = 0.4$, and $\alpha_\mathrm{F}  = 0.1$ are estimated according to \citet{Makwana2020PhRvX}.

Another uncertain parameter is the regular magnetic field. As particles of all energies are equally accelerated via TTD, the turbulent magnetic energy is consumed at the particle acceleration (i.e., energy gain of particles in scattering of waves is in turn a damping process for waves). The regular magnetic field, which does not decay with the wave damping, needs to be estimated. The typical magnetization parameter of ISM is $\sigma = B_{\mathrm{ISM}}^2/4\pi \rho c^2 \sim 10^{-9}$, giving a typical magnetic field of $B_{\mathrm{ISM}}\sim {\mathcal O}(\mu$G). The downstream magnetic field amplified by the shock compression is $4 \Gamma B_{\mathrm{ISM}}$ according to jump conditions of relativistic shock \citep{Blandford1976PhFl}, which provides the regular component of the downstream magnetic field.
Numerical simulations observe a magnetic equipartition $\varepsilon_B \sim 10^{-3}$ downstream of an unmagnetized collisionless shock \citep{Groselj2024ApJL}. This turbulent field, self-generated by current filamentation instabilities at the shock, decays slowly in the downstream region. As lower values of $\varepsilon_B$ have been inferred in some GRBs, which can be down to $10^{-5}$, and even $10^{-6}$ at late time, \citep{Panaitescu2002ApJ,Kumar2009MNRAS,Kumar2010MNRAS,BarniolDuran2011MNRAS}, this post-shock decay must be operative. Here we assume an upstream magnetic field  $B_{\mathrm{ISM}}= 5\,\mu$G, which is the typical ISM magnetic field. 

For parameters of the afterglow model, we follow the best-fitting results given by \citet{LHAASOCollaboration2023Sci} with a slight modification in order to adapt to the stochastic model. The important parameters are shown in \Cref{para}. In our model, the additional parameters are the dimensionless eddy scale $\xi$, the fraction $f_{\mathrm{ra}}$ of re-accelerated electrons, the downstream turbulence equipartition factor $\varepsilon_\mathrm{T}$, the magnetic component for turbulence $\alpha$ and for fast mode wave $\alpha_\mathrm{F}$. The latter two are fixed in the calculation. 
We align the above two FP equations \Cref{particle pde,turb pde}, and use the differential scheme developed by \cite{Chang1970JCoPh} to solve the evolution of particles and turbulence. The treatment of particle injection, escape and turbulence damping terms follows the procedure in \cite{Park1996ApJS}.

\section{Results and Discussion}\label{sec:Results and Discussion}

\subsection{Evolution of Particles and Turbulence\label{sec: particle spec}}
\begin{figure}[ht]
    \centering
    \includegraphics[width = \linewidth]{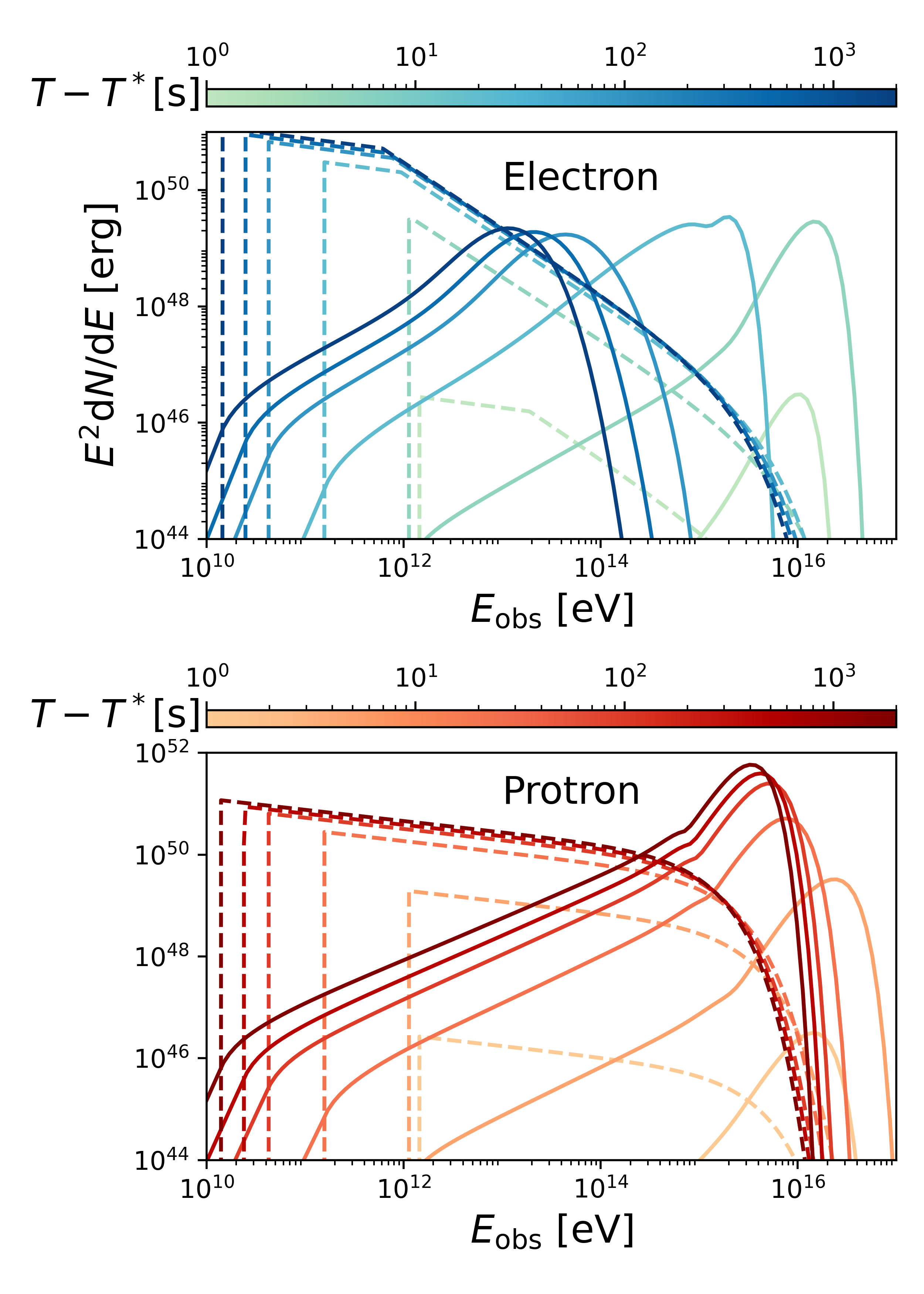}
    \caption{
        Evolution of the electron (Top Panel) and proton spectra (Bottom Panel). The solid lines represent the spectral energy distributions of SA particles, while the dashed lines indicate those of SH particles. The color bar refers to different times in the observer frame. 
        }
    \label{fig:particle spec}
\end{figure}

\Cref{fig:particle spec} illustrates the evolution of electrons and protons in the downstream turbulence.
The stochastic acceleration dominated by TTD predicts a hard particle spectrum ($p=1$). 
This indicates that most energy is concentrated in the high-energy end, forming a bump-like feature around the peak energy. 
At $T-T^* < 10$~s, the radiative cooling timescale of electrons is longer than the dynamical timescale even for those injected at the high-energy end of the spectrum.  
The energy spectra of electrons and protons then have the same peak energy, since they are subject to the same physical processes.
As time approaches $T-T^* \sim 10$~s, the inverse Compton cooling timescale of the high-energy electrons becomes comparable to the acceleration timescale, causing them to cool down. This results in a faster decrease of peak energy in the electron spectrum compared to the proton spectrum.

The peak energy shift could also be interpreted in terms of the decreasing acceleration efficiency. Due to the strong damping effect, the peak of turbulence occurs at the injection scale $\lambda = \xi R/\Gamma$. At a given energy in the comoving frame, the diffusion coefficient is proportional to
$D_{{EE}} \propto E^2(\delta B^2/B^2) (\Gamma/R)$, where $B^2 = B_0^2 + \delta B^2$ is the local magnetic field. 
So the acceleration timescale scales as $t_\mathrm{acc} \propto R/\eta\Gamma$
with $\eta = \delta B^2/B^2$ the relative magnitude of the turbulent magnetic density. 
The damped downstream magnetic field weakens synchrotron radiation, thus the energy loss of SA electrons is mainly due to inverse Compton scattering, with the cooling timescale $t_\mathrm{IC} \propto 1/ U_\mathrm{B,sh}\propto 1/{n\Gamma^2}$.
The ratio between the acceleration and cooling timescales is then $t_{\mathrm{acc}} / t_{\mathrm{IC}} \propto R\Gamma\eta^{-1}$. Besides the $R$ and $\Gamma$ determined by afterglow dynamical evolution, this also implies that increasing number of particles can lead to stronger damping, smaller $\eta$ and consequently a longer acceleration timescale compared with the cooling timescale.
The critical energy that balances electron acceleration and cooling shrinks to lower values, thus reducing the SA electron peak energy.

The peak energy of protons is determined by the resonant energy $E_{\mathrm{res}} = eB/k_{\mathrm{min}}$ of the minimum wave number. 
In the specific case of GRB~221009A, with a magnetic equipartition factor smaller than the typical value $0.025$ assumed in \cite{Zhang2021PhRvD}, and a small injection scale of turbulence ($\xi  = 10^{-2}$), the acceleration of UHE cosmic rays is found to be unfavorable.

The appearance of a bump-like structure around the peak energy region is another critical feature.
This feature is due to joint play of acceleration and processes that would prevent acceleration, such as the energy loss timescale or the dynamical timescale\footnote{Escape could also prevent particle acceleration, but it is not important in our case.}. Below the peak energy, the acceleration timescale is the shortest among all relevant timescales and particles are efficiently accelerated to higher energies. The cooling timescale or the dynamical timescale is shorter than the acceleration timescale above the peak energy. Particles cannot be accelerated to higher energies and therefore accumulate around the critical energy, resulting in the bump.
Interestingly, this phenomenon is not observed in \cite{Zhang2021PhRvD} with large $\xi$. 
In their gyro-resonance model, $t_\mathrm{acc}/t_\mathrm{dyn}$ changes slowly with energy. The resultant bump spans over larger energy range and is therefore not obvious.

\begin{figure}[ht]
    \centering
    \includegraphics[width=\linewidth]{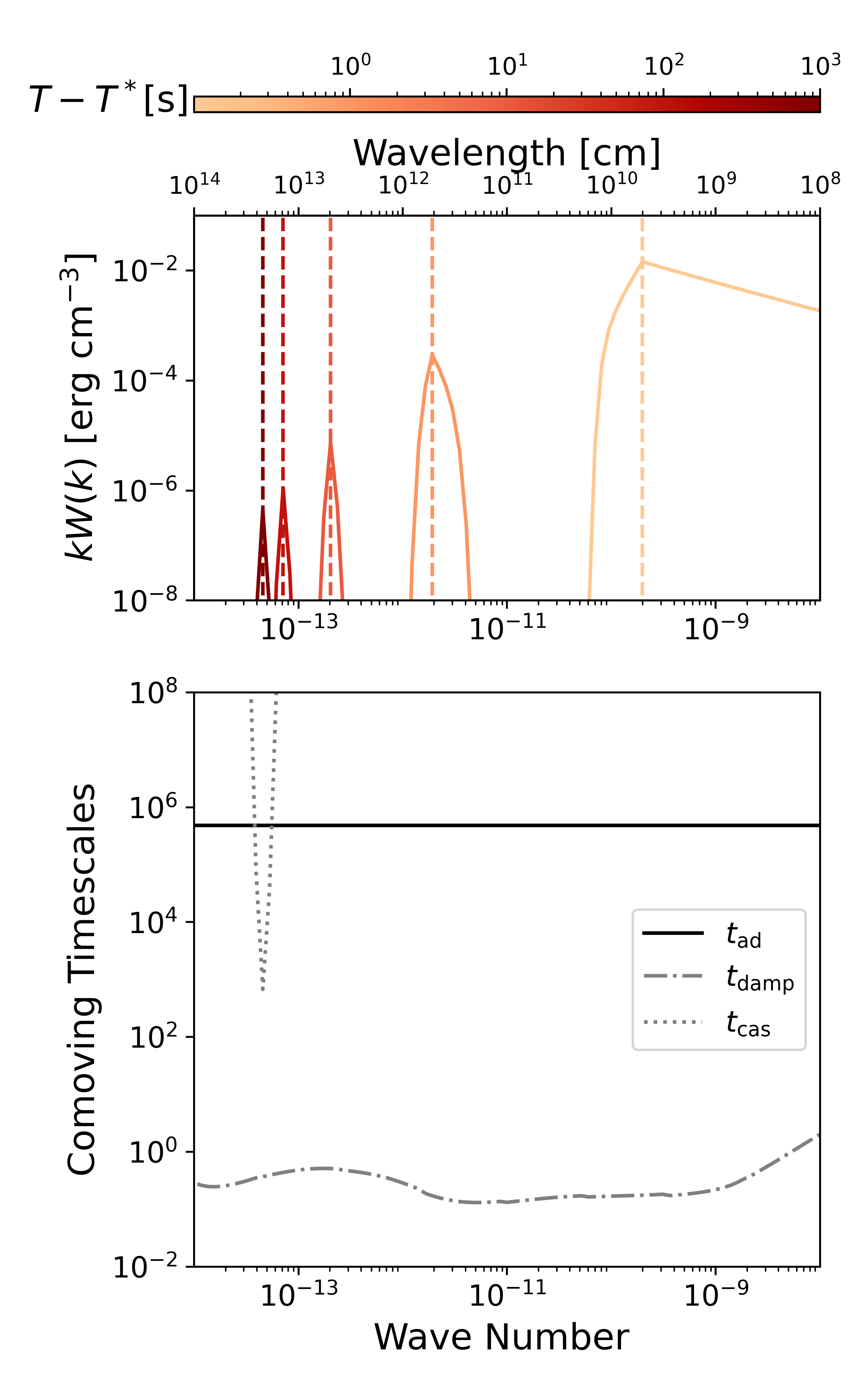}
    \caption{Top Panel: turbulent spectra at different times.
        The dashed lines indicate the wave number at which the turbulent energy is injected.
        Bottom Panel: dynamical timescale (Solid), cascade timescale (Dotted) and damping timescale (Dash-dotted). Timescales are plotted at $T-T^* = 1000$~s and measured in the comoving frame. For definitions, see \Cref{app:timescale}.}
    \label{fig:turbulence spec}
\end{figure}

The evolution of turbulence is shown in \Cref{fig:turbulence spec}. Since TTD is non-gyro-resonant, particles of all energies extract energy from the magnetic fluctuations. The scattering of particles leads to slowing down of escape, and hence trapping in the turbulent zone. As a result, damping is stronger as more particles are injected, so that eventually only the fresh injected turbulence can exist and resonate with particles. This is illustrated by the vertical dashed lines in \Cref{fig:turbulence spec}, which denotes the injection wave number at different times. Another consequence is that for particles of low rigidities ($r_\mathrm{g}k_\mathrm{min}\ll 1$), no turbulence is available to interact with particles via gyro-resonance, and the acceleration is completely dominated by TTD. Only at $r_\mathrm{g}k_\mathrm{min}\ll 1$ is the gyro-resonance significant with $D_{EE,\,\mathrm{gyr}} \propto E^0$, see e.g. \cite{Demidem2020PhRvD}. The decrease of injection wave number is due to the decelerating expansion of the blast, i.e. $k_{\mathrm{min}} \propto \Gamma/R$. Furthermore, as time progresses, the magnetic field predominantly consists of its regular component, with the amplitude scaling as $B_0 \propto \Gamma$.
The result of this trend is the spectral index evolution at VHE band, as we will interpret in the following section. 

\subsection{VHE Radiation\label{sec:VHEspec}}

\begin{figure}[ht]
    \centering
    \includegraphics[width=\linewidth]{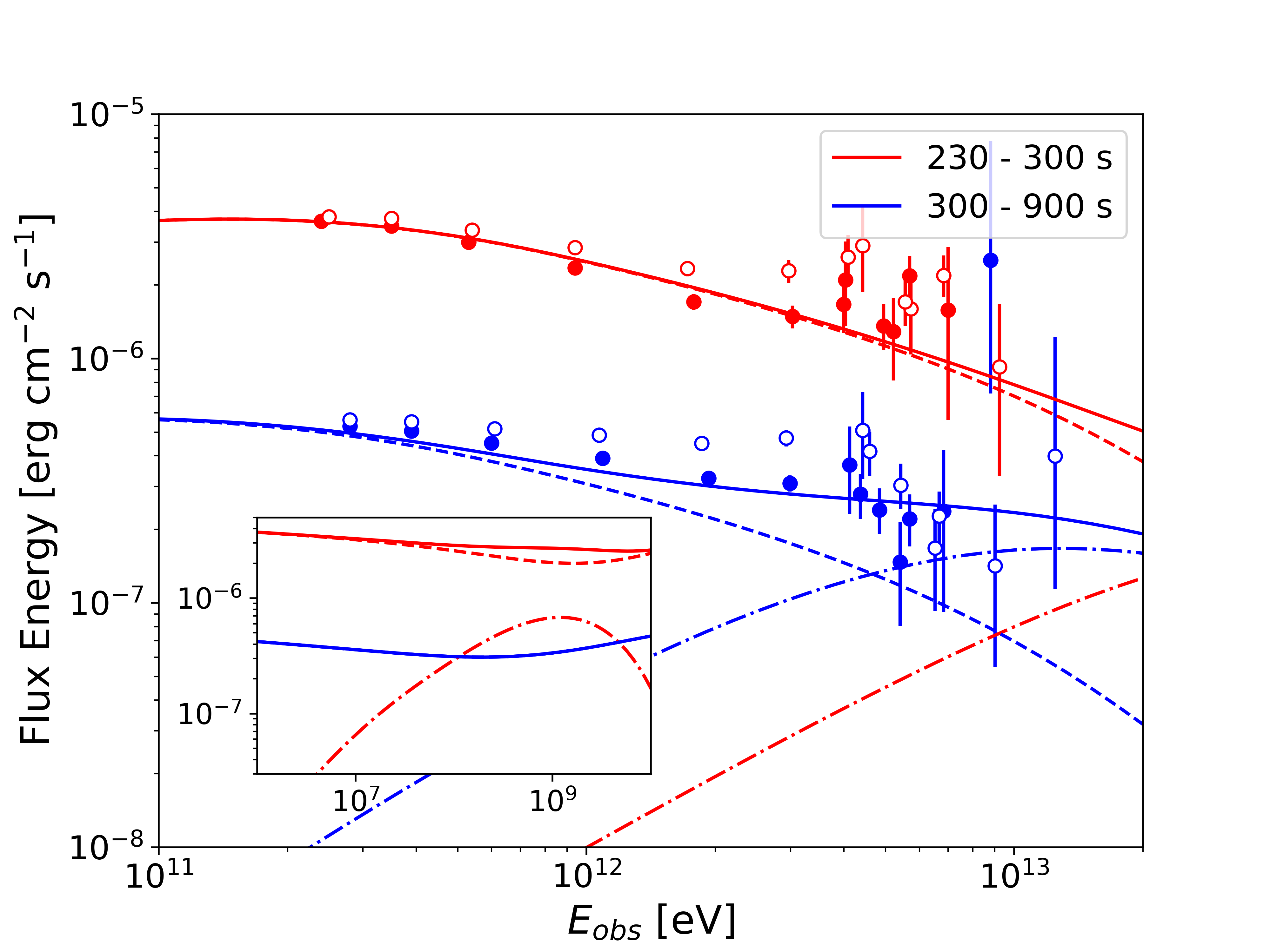}
    \caption{ The radiation spectrum in VHE band predicted by SA electrons and the standard afterglow model.
        The dashed lines show the radiation from SH electrons, while the dash-dotted lines correspond to SA electrons.
        The solid lines stands for total flux energy.
        The data are averaged over two time bins, $T_0 + [230,\,300]\,\mathrm{s}$ (red) and $T_0 + [300,\,900]\,\mathrm{s}$ (blue).
        The data points have been reduced following \citet{Cao2023SciA}; the filled and open circles represent the absorption-corrected  flux with EBL models of Saldana-Lopez 2021~\citep{Saldana-Lopez2021MNRAS} and Finke 2010~\citep{Finke2010ApJ}, respectively. SSC to first order and the intrinsic $\gamma\gamma$ absorption are considered.
        Inset displays the MeV to GeV radiation, showing that the synchrotron radiation of the SA component is not significant.}
    \label{fig:Radiation Spec}
\end{figure}

The gamma-ray spectrum is presented in \Cref{fig:Radiation Spec}. By averaging the radiation spectrum over two time bins ($T_0+[230,\,300]$~s and $T_0+[300,\,900]$~s), we obtain a consistent result that aligns with the findings of \cite{Cao2023SciA}. The shock SSC model alone (dashed line) fails to explain the hard spectra at $T_0 + [300,\,900]$~s, so adding the SA component is necessary. The final result agrees with the LHAASO KM2A observations, taking into account the radiation mechanisms mentioned in \Cref{radiation}. 
In particular, for $T < T_0 + 300$~s, the contribution from the turbulent electron component is found to be insignificant, because the peak energy in the electron spectrum is beyond 10\,PeV in the observer frame, resulting in a minor contribution at 10\,TeV. As the peak energy in the electron spectrum decreases with time, the peak of the IC spectrum produced by the turbulent electron component shifts towards lower energies, resulting in a higher flux around 10\,TeV.
The observed multi-TeV flux by LHAASO is attenuated by the extragalactic background light (EBL). \citet{Cao2023SciA} provided EBL-corrected spectrum with different EBL models. Here, we consider the spectrum corrected with the EBL model of \citet{Saldana-Lopez2021MNRAS} as the benchmark case (solid circles in \Cref{fig:Radiation Spec}), but also show the spectrum corrected with the EBL model of \citet{Finke2010ApJ} for comparison (open circles in \Cref{fig:Radiation Spec}).

\begin{figure}[ht]
    \centering
    \includegraphics[width=\linewidth]{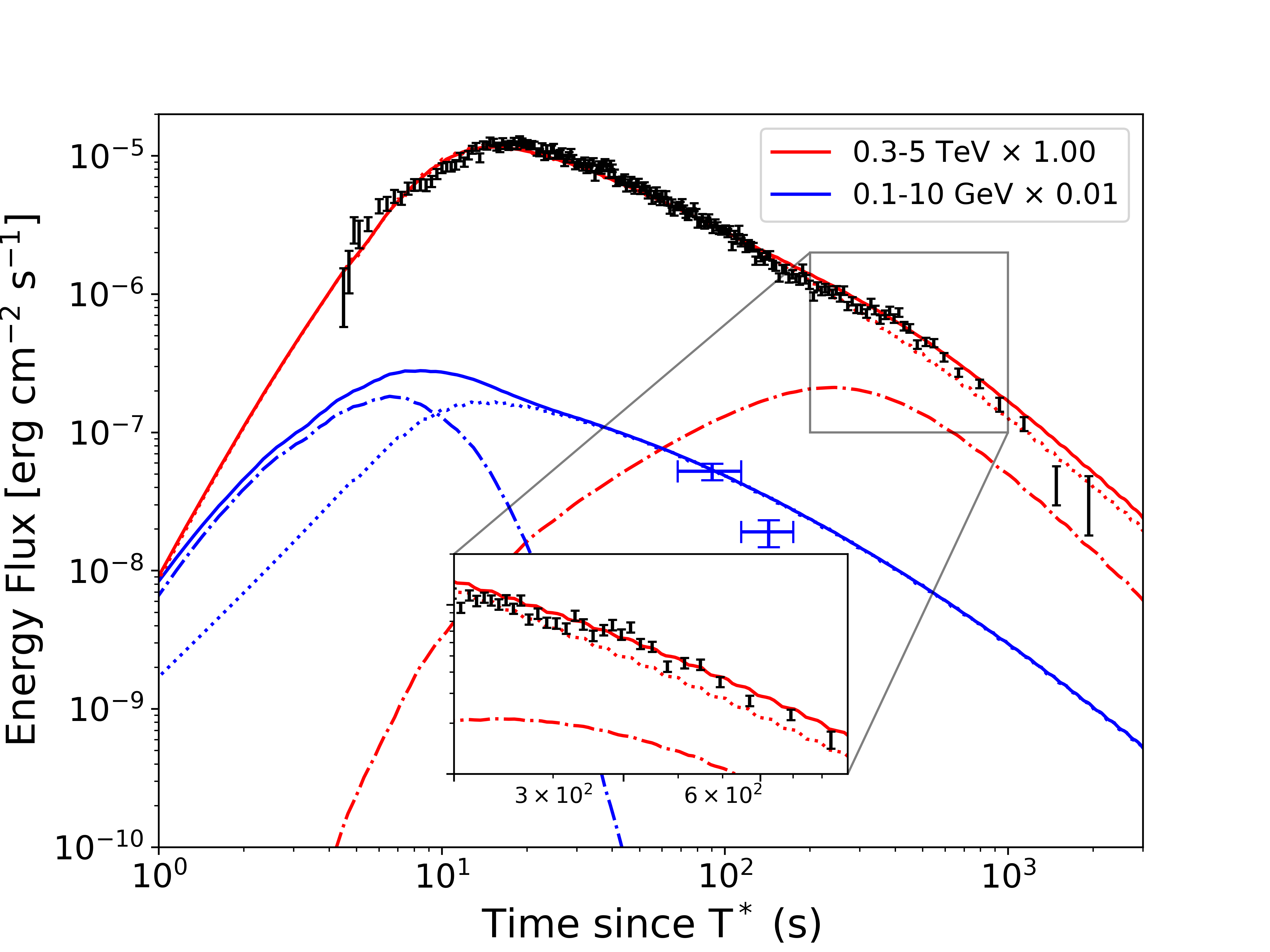}
    \caption{Light curve of GRB 221009A afterglow in three bands, 0.3-5 TeV (Red), 0.1-10 GeV (Blue).
        The dotted lines represent the radiation from SH electrons, the dash-dotted lines that from the SA electrons,
        and the solid lines the total.
        The data were measurements by LHAASO-WCDA \citep{LHAASOCollaboration2023Sci} (in black), and Fermi-LAT \citep{Liu2023ApJL} (in blue). In the keV band, there are Swift-XRT observations at $T^* + 10^4$~s, which are not displayed because the SA component does not contribute to the X-ray flux, and because the X-ray flux dominated by SH electrons is consistent with observation.
        The light curves are multiplied with different factors for clarity.}
    \label{fig:LC}
\end{figure}

Note that the last data point around
10\,TeV in the benchmark case shows an intriguing feature of steep hardening. Although stochastic acceleration predicts a hard electron spectrum, it is still not sufficient to explain this steep hardening. Indeed, 
the low-energy tail of the IC scattering of mono-energetic electrons can give rise to the hardest possible spectrum, characterized by $dn/d\varepsilon \propto \varepsilon^0$ (reference), while the data indicates $dn/d\varepsilon \propto \varepsilon^3$ approximately. However, the uncertainty of the measured flux is large, and other EBL models may predict a different opacity (see open circles). Therefore, we do not aim to reproduce the steep hardening around 10\,TeV.

\subsection{Multi-Wavelength Light Curve}
Since SA electrons present a hard spectrum that peaks at energies greater than 10\,TeV within $\sim T^*+1000$ s, their radiation mainly concentrates on the GeV band through synchrotron radiation and the TeV band via IC scattering, and does not affect the lower energy light curves. As shown in \Cref{fig:LC}, an early GeV excess arises from SA electrons, through synchrotron radiation of electrons accelerated to PeV energies. 
This component could also be significant for other GRBs with large downstream magnetic equipartition factors and might, in principle, be observable by Fermi-LAT.
The contribution of SA electrons to the TeV light curve becomes non-negligible since $\sim T^*+100$~s, when the peak energy in spectrum of SA electrons shifts below $\sim 100$\,TeV in the observer's frame and the peak of the corresponding IC spectrum shifts into the range of $0.3-5\,$TeV. At later times, the IC flux of SA electrons decreases because density of the target radiation field becomes smaller with the expansion of the shock.

\begin{figure}[ht]

    \centering
    \includegraphics[width=\linewidth]{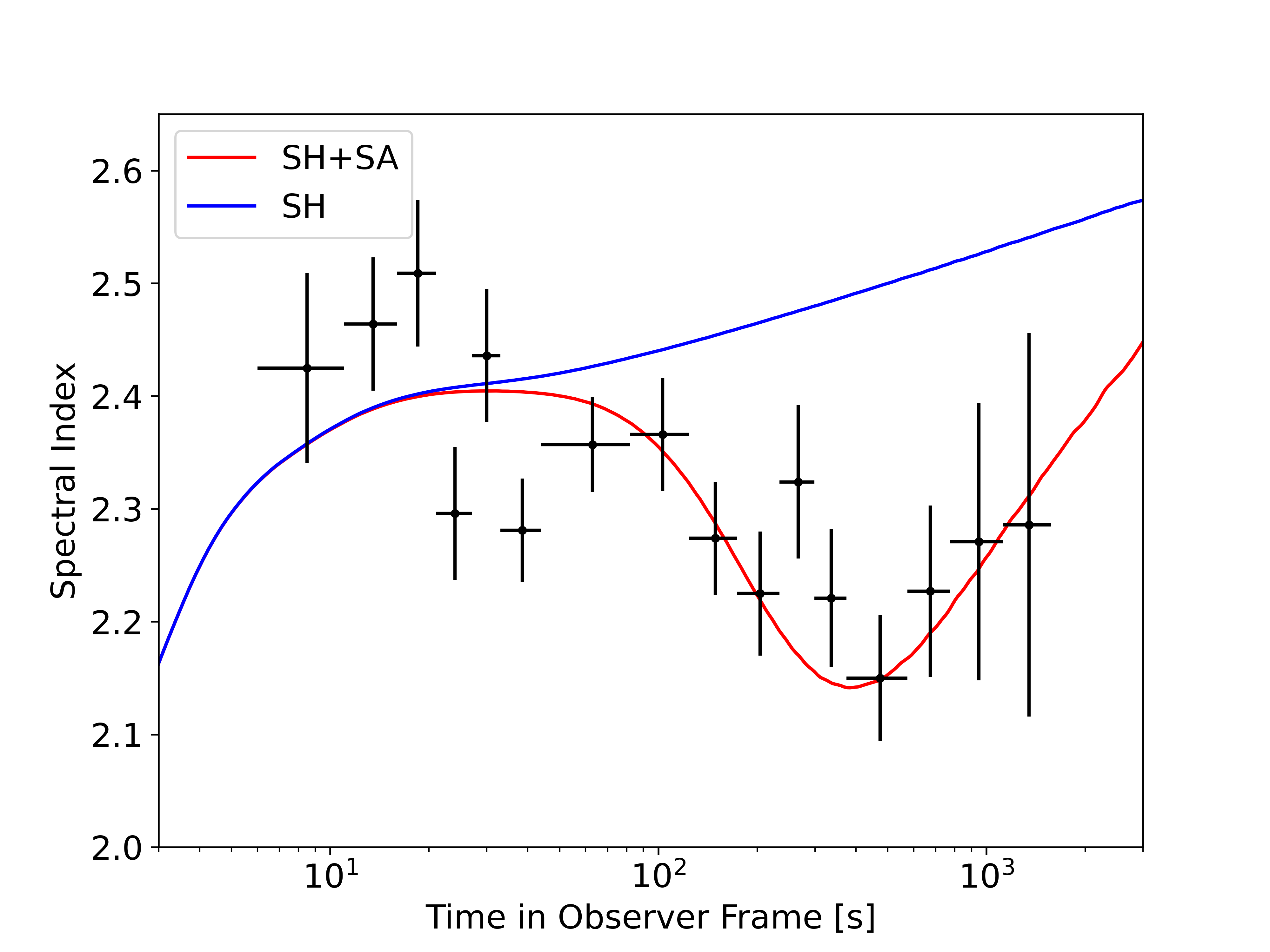}
    \caption{Spectral index of radiation.
        The data points are from LHAASO-WCDA \citep{LHAASOCollaboration2023Sci} at 0.3--5\,TeV.
        The solid lines represent the model prediction, fitted with a power-law model.
        Red: SH and SA electrons; Blue: SH electrons only.
        }
    \label{fig:Spec Index}
\end{figure}

The fitting to the temporal evolution of the spectral index of the VHE spectrum within $0.3-5$\,TeV \citep{LHAASOCollaboration2023Sci} is shown in \Cref{fig:Spec Index}. The standard afterglow model (i.e, the SSC radiation of SH electrons), denoted by the blue solid line, predicts a break in the evolution of the spectral index around the shock deceleration time (i.e., $T^* + 10\,\text{s}$), but the overall evolution presents a monotonic softening. This is inconsistent with the observation.
The addition of SA electrons significantly improves the fitting to the evolution of the spectral index, as shown by the red solid line. As the contribution of the SA electrons becomes important, the total spectrum is hardened from $\sim T^*+100$\,s. The hardest spectrum of our model is around $T^* + 353$\,s, with spectral index $\gamma = 2.14$. At later time, the spectral peak of SA electrons continuously shifts towards lower energies, and the total spectrum becomes soft again. Our model prediction is consistent with the LHAASO data, although the second break of softening in the spectral index evolution is not confirmed observationally.

\subsection{Implication for other GRBs}

The external shock of GRB~221009A does not show specific properties with respect to other GRBs, except the narrow jet derived from the early break in the TeV light curve. We therefore speculate that stochastic acceleration may operate in other GRB afterglows as well. The relative importance of the radiation of SA electrons are dependent on parameters of afterglows, which could vary from burst to burst. 

One important parameter for SA electrons is the magnetic equipartition coefficient. A larger $\varepsilon_B$ leads to a higher magnetic field strength and also more energy in the turbulence. This would lead to a more prominent GeV excess at early time as well as the TeV bump at later time. In particular, if $\varepsilon_B$ is sufficiently high, the VHE radiation of the SA component could exceed that of SH electrons at lower energies (i.e. $\lesssim$TeV), which is easier to be observed because of less EBL attenuation.  

Another important yet uncertain parameter is the turbulence injection scale $\xi$.  For a larger injection scale of turbulence, the corresponding particle energy at resonance  (i.e. $E_{\mathrm{res}} (k_{\mathrm{inj}})$) also becomes higher. As we mentioned earlier (see \Cref{fig:particle spec}), the peak energy of proton spectrum is higher than that of electron spectrum because of the much more efficient cooling of electrons. Therefore, as the turbulence cascades down to smaller scales from the injection scale, the turbulence energy will be tapped and consumed by protons before it can efficiently resonate with electrons. Consequently, the electron acceleration will be suppressed, but protons can be accelerated to higher energies, which is similar to the scenario considered by \citet{Zhang2021PhRvD}. 

It may be also worth mentioning that the discussion above is based on the assumption of ISM as the circumburst medium. If the circumburst environment is dominated by the stellar wind of the progenitor star, the result will be slightly different. Similar to the ISM scenario, the electron energy loss is dominated by inverse Compton scattering off the synchrotron radiation by SH electrons.
As the number density of the circumburst medium $n \propto 1/R^2$,  we then have $t_\mathrm{IC}/t_\mathrm{acc} \propto R \eta/\Gamma$, which grows monotonically as time progresses (\Cref{app:windenv}). Therefore, the SA electron peak energy will increase. Far away from the central engine, the decreasing radiation energy loss rate enables electrons to be accelerated to higher energy. As a result, the peak energy of SA electrons at early times will be relatively low, then shifts towards higher energies, and eventually contribute most of their radiation energy above 10\,TeV, producing a hard VHE spectral component.

\subsection{Parameter Degeneracy}
In this work, we manually adjusted the parameters $\varepsilon_B$, $\xi$ and $f_\mathrm{ra}$ to reproduce the observed afterglow emission. Although the parameter settings shown in Table \ref{para} match observations in multiple wavelengths, they are not exclusive and certain degree of degeneracy exists in the parameter space. 

The main degeneracy lies between 
the turbulence injection scale $\xi$ and the magnetic equipartition coefficient  $\varepsilon_{\rm B}$. For a larger turbulence injection scale or a larger $\xi$, the fraction of turbulence energy which is cascaded to the gyro-scale of electrons with relevant energies (TeV) becomes smaller, resulting in a less efficient electron acceleration. In this case, we need increase $\varepsilon_B$ (or the total turbulence energy at injection) accordingly, in order to keep the acceleration efficient and reproduce LHAASO observations. A higher $\varepsilon_B$ will result in stronger synchrotron radiation of the SA component. At early time up to $\sim T^* + 10$~s, however, the more efficient synchrotron cooling will not hinder the particle acceleration since the dynamical timescale is shorter than the cooling timescale at early time. Without efficient cooling, electrons and protons are accelerated to equipartition energies. The peak energy and the flux of the synchrotron radiation would thus be higher for a higher $\varepsilon_{\rm B}$. As a result, the degeneracy may be resolved with early GeV gamma-ray observations. Unfortunately, the observation of GRB~221009A before $T^*+10$~s with Fermi-LAT suffers pile-up effect \citep{Axelsson2025ApJS}, we therefore can not use this observation to constrain the parameters. Moreover, the turbulence injection scale may be estimated via numerical simulations of relativistic shock, which is deferred to future studies.

\section{Summary}\label{sec:summary}

We explored the stochastic acceleration model in a GRB afterglow with a Fokker-Planck approach, and successfully explained the LHAASO observation of GRB~221009A afterglow. Besides GRB~221009A, we have discussed the possible observable features resulted from SA electrons in other gamma-ray bursts that could be used to verify our model. Our main conclusions are summarized as follows:

\begin{itemize}[itemsep=1ex]
    \item The hard spectrum of the afterglow of GRB~2210009A observed by LHAASO could be explained by adding a second electron component arising from the stochastic acceleration in the downstream turbulence. The SA electron hardens the VHE spectrum after $T^* + 100$~s, consistent with the observation.
    
    \item By adding the SA component, we improve the fitting to the spectral index evolution measured by LHAASO significantly, compared to the standard afterglow model. The SA component vanishes after several thousands of seconds due to less and less efficient acceleration. 
    
    \item SA electrons may also lead to an excess in the GeV band at early time, produced by the synchrotron radiation. The GeV excess may be more prominent in those GRBs with larger $\varepsilon_B$.
\end{itemize}

In our calculation, we assumed that SA electrons cannot return to the shock. In other words, evolution of SH electrons and SA electrons are calculated in two separate single zones, except that radiation of SH electrons serves as targets for IC scattering of SA electrons. In reality, there may be exchange between SA electrons and SH electrons. Whether the interplay between SH and SA electrons is important need be explored in numerical studies with consideration of an additional space dimension in the model. 
Additionally, we have based our study on acceleration via resonant wave-particle interactions, but note that non-resonant acceleration mechanisms, in particular a generalized Fermi process \citep{Lemoine2021PhRvD,Lemoine2022PhRvL} can provide comparable, if not dominant contributions in the strong turbulence limit. We defer the examination of these effects to a future study.

\vspace{1em}

\section*{Acknowledgements}
We express gratitude for the valuable discussions with Daniel Groselj and the support from Hua-Feng Yu. This work is supported by the National Science Foundation of China under grants No.12393852 and No.12333006.

\appendix
\section{Model Detail\label{app:model}}
\subsection{Resonance Kernel\label{app:kernel}}

The resonance kernel in \cite{Lemoine2024PhRvD} can be expressed as
\begin{equation}
    \Phi \propto \left[(r_\mathrm{g}  k)^{1/\Delta_1 }+ (r_\mathrm{g}  k)^{-1/\Delta_2}\right]^{-1}\text{,}
\end{equation}
The normalization of $\Phi(k,E)$ is defined by $\int \Phi(r_\mathrm{g},k) \,\mathrm{d}k \equiv 1$.
There are two parameters $\Delta_1$ and $\Delta_2$, with $r_\mathrm{g} = E/eB$ the gyro-radius.
This function shows the interaction strength of particle with given energy with wave of different $k$-modes.
This is a power-law function with different slope $-1/\Delta_1$ and $1/\Delta_2$ at each side of the resonant wave number.

Particle acceleration is dominated by turbulence modes with $r_\mathrm{g}  k<1$,  thus the parameter $\Delta_1$ is unimportant as long as  it is not larger than $1$. We simply set it as $\Delta_1 = 0.5$, so that $D_{EE}$ scales as $E^{-1}$ when $r_\mathrm{g}  k<1$, same as the analytical form of TTD for fast mode \cite{Demidem2020PhRvD}. 

According to QLT prediction, $D_{pp}$ scales as 
\begin{equation}
    \int g(k) k^{-1} J_1^2(z) \, \mathrm{d} k \propto \int k^{1-q} \, \mathrm{d} k\propto \int \Phi k^{-q}\, \mathrm{d}k
\end{equation}
for TTD, where $g(k) \propto k^{-q}$ is the power spectrum of magnetic fluctuation \citep{Jaekel1992JPhG}. This result requires $\Delta_2 = 1$. With this value, $D_{EE}$ scales as $E^q$ for $E_\mathrm{res}   (k_\mathrm{max}  )<E < E_\mathrm{res}   (k_\mathrm{min}  )$, and $E^2$ for $E < E_\mathrm{res}   (k_\mathrm{max}  )$. The corresponding particle spectral index should then be $p =1-q$ and $p= -1$.

The analytical form of $D_{EE}$ in previous works \citep{Teraki2019ApJ,Demidem2020PhRvD} assumes undamped turbulence, and yields
\begin{equation}
    D_{\gamma\gamma} \sim \gamma^2\beta_\mathrm{ph}  ^2 c k_\mathrm{res}  ^2 \frac{P_B(k_\mathrm{res}  )}{B_0^2} f_\mathrm{res}  \text{,}\label{eq:dee_teraki}
\end{equation}
where $\gamma$ is the particle Lorentz factor; $\beta_\mathrm{ph}   = v_\mathrm{ph}  /c$ is the dimensionless phase velocity of fast mode magnetosonic wave, which is defined by \citep{Zhang2019pgrb}
$$\beta_\mathrm{ph} = \frac{v_\mathrm{ph}}{c} = \sqrt{\frac{\hat{\gamma}P + B^2/4\pi}{\rho c^2 + \hat{\gamma}P/(\hat{\gamma}-1) + B^2/4\pi}}\text{,}$$
with $\hat{\gamma} = 4/3$ the relativistic adiabatic index, $P = 4\Gamma^2 n_\mathrm{ISM} m_\mathrm{p} c^2/3$ the pressure, and $\rho = 4\Gamma n_\mathrm{ISM}m_\mathrm{p}$ is the downstream rest mass density. 
$k_\mathrm{res}   = eB/E$ is the resonant wave number for particles with given energy; $f_\mathrm{res}  $ denotes the fraction of particles in TTD resonance with respect of resonance broadening.

In our case, where damping is significant, the energy distribution of turbulence will deviate from the initial spectrum.
Thus we adopt the phenomenological kernel description of wave-particle interaction \citep{Lemoine2024PhRvD}. Since we assume strong turbulence, and the fast mode phase velocity is relativistic, we adopt the normalization from \cite{Teraki2019ApJ} and let $f_\mathrm{res}=1$.

\subsection{Evolution Timescales\label{app:timescale}}

Radiation energy loss timescale is computed as $t_\mathrm{rad} \equiv E/P_\mathrm{rad}$, with $P_\mathrm{rad}$ the power of radiation. 

Wave damping timescale is $t_\mathrm{damp} = 1/\Gamma_W$.
The cascade timescale takes the form $t_{\mathrm{cas}} \equiv k^2 / D_{kk}$, i.e.
\begin{equation}  
    t_{\mathrm{cas}} = \left\{\begin{aligned}
         & \frac{1}{k\beta_{\mathrm{sh}}c}\left[\frac{2U_B}{kW(k)}\right]^{1/2} & \mathrm{(K41)} \\
         & \frac{1}{k\beta_{\mathrm{sh}}c}\left[\frac{2U_B}{kW(k)}\right]       & \mathrm{(IK)}
    \end{aligned}\right. \text{.}
\end{equation}

\subsection{Damping\label{app:damp}}

To derive the damping term $\Gamma_W(k)W(k)$, we consider the contribution to particle acceleration and damping of a narrow range of turbulence in wave number space between $k$ and $k+\delta k$.
For TTD, the turbulence at wave number $k$ leads to the particle diffusion \begin{equation*}
    D_{{EE},k} = \frac{E^2 \beta_\mathrm{ph}  ^2 c k_\mathrm{res}  ^2}{U_B}\Phi(E,k)W_\mathrm{F} (k) \delta k\text{.}
\end{equation*}

We use
\begin{equation}
    F_\mathrm{p}(E) = E^2 D_{{EE},k} \frac{\partial}{\partial E}\left(\frac{N}{E^2}\right)\label{fp}
\end{equation}
to rewrite the first two terms on r.h.s. in \Cref{particle pde}, ignoring particle injection, escape and energy loss, \Cref{particle pde} becomes
\[\frac{\partial N(E) }{\partial t} = \frac{\partial F_\mathrm{p}(E)}{\partial E}\text{.}\]

Energy conservation requires
\begin{equation}
    - \Gamma_{W,\rm TTD}(k)W(k)\delta k = \int E\frac{\partial n(E)}{\partial t}\,\rm{d}E \label{econserv}\text{,}
\end{equation}
where $n = N/ (4\pi R^3/\Gamma)$ is the particle number density in the SA region, and the time derivative $\partial n/\partial t$ denotes that the particles are only accelerated by turbulence at $k$.
Substituting \Cref{fp} into  \Cref{econserv} and exchange the order of integral, we get
\begin{equation}
    \begin{split}
        \Gamma_{W,\rm TTD}(k)
        =-{8\pi e^2 \beta_\mathrm{ph}  ^2 c \alpha_\mathrm{F} }
        \int_\mathrm{e} dE\left[E^2 \frac{\partial }{\partial E}\left(\frac{n}{E^2}\right)\frac{\partial \Phi(k,E)}{\partial E} \right. 
            +\left. \frac{\partial}{\partial E}\left(E^2\frac{\partial }{\partial E}\left(\frac{n}{E^2}\right)\right)\Phi(k,E)\right]\text{.}
    \end{split}
\end{equation}

Similarly, the damping coefficient for gyro-resonance is \citep{Zhang2021PhRvD}
\begin{equation}
    \Gamma_{W,\,\rm gyr}(k) = -\frac{8\pi e^2 \beta_\mathrm{ph}  ^2 c \alpha_\mathrm{F} }{k} \left[n(E_\mathrm{res}) + \int _{E_\mathrm{res}} ^{E_\mathrm{max}}\frac{2n(E_\mathrm{res})}{E}\, \mathrm{d}k\right]
\text{.}
\end{equation}

\subsection{Radiation\label{app:radiation}}
\subsubsection{Energy Loss}
We adopt the simplified energy loss rate from \cite{Schlickeiser2010NJPh}. 

\begin{equation}
   b'(\gamma ',t',r_i) = \frac{4}{3}\sigma _\mathrm{T} \frac{1}{m_\mathrm{e}c}\left[U_B(r_i) + F_\mathrm{KN} (\gamma ',t',r_i) u_\mathrm{soft}'(t',r_i)	\right]\label{eq:lossrate}\text{,}
\end{equation}
where $\sigma_T$ is the Thomson cross section, $u_B'(r_i ) = B^2(r_i)/8\pi$
is the magnetic energy density.
\begin{equation}
   F_\mathrm{KN} (\gamma ',t',r_i) = \frac{9}{u'_\mathrm{soft}(t',r_i)}\int_{0}^{\infty} \, \mathrm{d}\epsilon \epsilon n'_\mathrm{soft}(\epsilon,t',r_i) \times \int_{0}^{1} \, \mathrm{d}q\frac{2q^2\ln q+ q(1+2q)(1-q) + \frac{q(\omega q)^2(1-q)}{2(1+\omega q)}}{(1+\omega q)^3} \text{,}
\end{equation}
where $\omega = 4\epsilon \gamma' / (m_\mathrm{e}c ^2)$.

In \Cref{eq:lossrate}, $b$ is the energy loss rate.
The energy loss is then $\dot{\gamma} = b\gamma^2$.
Energy loss of the single electron in the magnetic field and soft target photon field is then $\dot{E} = b\gamma^2E_\mathrm{e}$.

\subsubsection{Intrinsic $\gamma\gamma$ Absorption }

The optical depth of $\gamma\gamma$-absorption is \citep{Jauch1976tper}

\begin{equation}
   \tau_{\gamma\gamma}(\varepsilon'_\gamma,\varepsilon) =\int_{\varepsilon_\mathrm{th}}\sigma_{\gamma\gamma}(\varepsilon'_\gamma,\varepsilon) \frac{R}{\Gamma}\frac{\mathrm{d}n'}{\mathrm{d}\varepsilon'} \, \mathrm{d}\varepsilon'\, \mathrm{d}\Omega
\end{equation}

\begin{equation}
   \sigma_{\gamma\gamma}(\varepsilon'_\gamma,\varepsilon) = \frac{3\sigma_\mathrm{T}}{16}(1-\beta^2)\times \left[2\beta(\beta^2-2)+(3-\beta^4)\log\left(\frac{1+\beta}{1-\beta}\right)\right]
\end{equation}

where $\beta = \sqrt{1-\varepsilon_\mathrm{th}/\varepsilon}$. $\varepsilon_\mathrm{th}$ is the threshold energy of $\gamma\gamma$ absorption is $\varepsilon_\mathrm{th}= 2m_\mathrm{e}^2c^4/\varepsilon_\gamma(1-\cos \theta)$.
The $\theta$ denotes the angle between the two photons.
In the optimal case, we have $\varepsilon_\mathrm{th} = 2 m_\mathrm{e}^2 c^4 / \varepsilon_\gamma $.

The flux after $\gamma\gamma$ correction is written as $F'(E) = F(E)e^{-\tau}$.

\subsection{ Demonstration of Monotonicity of $t_\mathrm{IC}/t_\mathrm{acc}$ in wind environment \label{app:windenv}}
In the discussion of SA in stellar wind dominated circumburst environment, we have shown that $t_\mathrm{IC}/t_\mathrm{acc} \propto R \eta/\Gamma$, where $\eta \equiv \delta B^2 / B_0^2$. 
Here we have assumed that at late time, with large total number of particles trapped in the downstream turbulence and low upstream number density, the downstream magnetic field is dominated by the regular component $U_{B0}$.   

Obviously, $R/\Gamma$  grows monotonically in the expression of $t_\mathrm{IC}/t_\mathrm{acc} $ above, while $\eta$ may decrease. We rewrite $\eta$ as  $\int W_B(k)\, \mathrm{d}k/U_{B0}$, where $U_{B0} = 16\Gamma^2 \sigma n_\mathrm{wind}m_pc^2/8\pi\propto \Gamma^2/R^2$, with the assumption of a radius-independent magnetization parameter $\sigma$. The $\int W_B(k)\, \mathrm{d}k $ on the numerator could be approximated by $\alpha Q_\mathrm{w,inj} t_\mathrm{damp}$, which represents the fresh injected turbulent magnetic energy. From Section 2, we have $Q_\mathrm{w,inj} \propto \Gamma^3/R^3$. The damping timescale relies on the inverse of particle number density, $1/n \propto (4\pi R^3/\Gamma)/N \propto R^2/\Gamma$. If we treat $t_\mathrm{damp}$ as a constant and compute a new $\eta' \propto \Gamma/R$, it drops more quickly than the actual $\eta$.

Calculating $(t_\mathrm{IC}/t_\mathrm{acc})_{\eta'}$ with $\eta'$, we yield $(t_\mathrm{IC}/t_\mathrm{acc})_{\eta'} = \rm{Const}.$, showing that $(t_\mathrm{IC}/t_\mathrm{acc})_{\eta}$ monotonically increases.

\bibliography{ref}
\bibliographystyle{aasjournal}

\end{document}